\documentclass{aa}
\usepackage{aabib99}
\usepackage{psfig}
\usepackage{psfrag}
\usepackage{amssymb}
\usepackage[dvips]{graphicx}
\usepackage{afterpage}

\title{On mesogranulation, network formation and supergranulation}

\author{M. Rieutord\inst{1,2}, T. Roudier\inst{1}, J.M. Malherbe\inst{3}
and F. Rincon\inst{4}}

\date{Received \today  / Accepted }

\offprints{M. Rieutord}

\institute{
Laboratoire d'Astrophysique de Toulouse,
Observatoire Midi-Pyr\'en\'ees, 14 avenue E. Belin, 31400 Toulouse,
France \and Institut Universitaire de France
\and DASOP, Observatoire de Paris,
Section de Meudon, 92195 Meudon, France
\and Ecole Normale Sup\'erieure de Lyon, 46 allee d'Italie, 69364 Lyon
Cedex 07, France}

\newcommand{\MG}{mesogranulation}
\newcommand{\SG}{supergranulation}
\newcommand{\ie}{{\it i.e.}\ }
\newcommand{\demi}{\frac{1}{2}}
\newcommand{\moy}[1]{\left\langle #1\right\rangle}
\newcommand{\lp}{ \left(}
\newcommand{\rp}{ \right)}
\newcommand{\cth}{ \cos\theta }
\newcommand{\sth}{ \sin\theta }
\newcommand{\noi}{ \noindent }
\newcommand{\vr}{\vec{r}}
\newcommand{\vv}{\vec{v}}
\newcommand{\en}{\vec{e}_n}
\thesaurus{
02.03.3, % Convection
02.20.1; % Turbulence
06.07.2, % Sun: granulation
06.16.2; % Sun: photosphere
}

\begin{document}

\authorrunning{Rieutord et al.}
\maketitle

\begin{abstract}
We present arguments which show that in all likelihood mesogranulation
is not a true scale of solar convection but the combination of the
effects of both highly energetic granules, which give birth to strong
positive divergences (SPDs) among which we find exploders, and averaging
effects of data processing.  The important role played by SPDs in
horizontal velocity fields appears in the spectra of these fields where
the scale $\sim$4~Mm is most energetic; we illustrate the effect of
averaging with a one-dimensional toy model which shows how two
independent non-moving (but evolving) structures can be transformed
into a single moving structure when time and space resolution are
degraded.

The role of SPDs in the formation  of the photospheric network is shown
by computing the advection of floating corks by the granular flow. The
coincidence of the network bright points distribution and that of the
corks is remarkable.  We conclude with the possibility that
supergranulation is not a proper scale of convection but the result of
a large-scale instability of the granular flow, which manifests itself
through a correlation of the flows generated by SPDs.

\end{abstract}

\section{Introduction }

In the traditional view of solar convection seen at the sun's surface,
three scales play the main roles: {\it granulation} (1~Mm) which shows
up as an intensity pattern most probably first seen by Herschel in 1801
\cite{BLD84},
{\it supergranulation} ($15 - 30$~Mm) which appears in (but not only)
dopplergrams of the full disk of the sun as a pattern of horizontal
velocities and which was first noticed by
Hart\cite*{Hart56b}\nocite{Hart56a} and confirmed by Leighton et al.
\cite*{LNS62}, and {\it mesogranulation} ($3 - 10$~Mm) observed by
November et al. \cite*{NTGS81} on Doppler measurements of vertical
velocities.

While the dynamics of granulation is rather well understood, its scale being
controlled by the balance of radiative diffusion of heat and convection, the
origin of the two other scales remains largely mysterious. The
ionization of helium was often invoked to explain these two scales since
the first and second ionizations of this atom occur at depths similar to
the mesogranular and supergranular scale respectively. However, this
geometrical  explanation is based on a ``laminar view" of solar
convection which is, on the contrary, very strongly turbulent.

The aim of the present paper is to investigate the dynamics of scales
larger than that of the granulation, using the horizontal flows given by 
granular motions determined by the new algorithms
described in Roudier et al. (1999). Briefly, these algorithms, Local
Correlation Tracking on binarized images (LCT$_{\rm bin}$) or Coherent
Structure Tracking (CST), allow us to increase noticeably the spatial and
temporal resolutions of the surface velocity fields; typically, we can
bring the spatial grid size down to 0\arcsec7 and the time step down to
5~mn.

Thus, we first concentrate on \MG\ (Sect.~2) and show the major role
played at this scale by strong positive divergences (SPDs) and by
averaging procedures.  It turns out, indeed,  that what has been
described in previous work as \MG\ results from a combination of a
physical phenomenon (SPDs among which are found exploding granules) and
a data processing effect applied to a turbulent flow (averaging). Since
each author had his own technique for averaging data, results have been
rather confusing and no clear-cut description of \MG\ has emerged. We
show here that when averaging is properly controlled, no quasi-steady
flow can be detected in the mesoscale range.

We then proceed (Sect.~3) with the investigation of the transport
properties of the mesoscale flows and show that the \SG\ scale appears
when the positions of concentrations of corks are compared to the
positions of network bright points. We conclude the
paper with a discussion of a model which seems to explain many of the
observations and the interactions of the three scales.

\section{Mesogranulation}

%\pagebreak
%\restylefloat{table}

\afterpage{\clearpage
\begin{table*}[h] \centering \vspace{1mm}
\begin{flushleft}\begin{tabular}{cccccccc} \hline

Year &  Reference & Spatial    & Time    & Field of &  Temporal & Spatial   & Measured \\
      &             & Resolution & step    & view     &  average  & average   & parameter\\
\hline
 1981 & November    &    1\arcsec & 85 sec & 60\arcsec$\times$160\arcsec    &  60 min.  & 1\arcsec $\times$ 1\arcsec    & Doppler  \\
      & et al. \cite*{NTGS81}&   &        &            &           & 3\arcsec $\times$ 3\arcsec    & velocity \\
      &             &            &        &            &           & 9\arcsec $\times$ 9\arcsec    &          \\
 \hline
 1982 & November    &    1\arcsec       & 85 sec & 166\arcsec$\times$140\arcsec   &  60 min.  & 3\arcsec $\times$ 3\arcsec    & Doppler  \\
      & et al. \cite*{NTGS82}&   &        &            &           &20\arcsec $\times$ 20\arcsec   & velocity \\
  \hline
1984  &Oda \cite*{Oda84}&  0.25\arcsec&  30 sec &  54\arcsec$\times$52\arcsec&  & &
Active  granules \\
      &  &  &  &  & & & Intensity \\ 
  \hline
 1986 &Koutchmy and & 0.5\arcsec  a 1\arcsec   & 15 sec & 90\arcsec$\times$67\arcsec     &  7.5 min. & Defocus   & Intensity\\
      &Lebecq \cite*{KL86}&      &42.8sec &  206\arcsec       &  46 min.  &   10\arcsec      &          \\
  \hline
 1987 & November    &    1\arcsec       & 10 sec & 166\arcsec$\times$250\arcsec   &  27 min.  & 4.2\arcsec$\times$4.2\arcsec  & Intensity\\
      & et al.\cite*{NSTTF87}&   &        &            &           &           &          \\
  \hline
 1987 & Dame and    &    1\arcsec       & 12 sec & 90\arcsec$\times$90\arcsec     & analyse   &  20 min.  & Intensity\\
      & Martic \cite*{DM87}&     &        &            & spectrale &           &          \\
  \hline
% 1988 & November    &    1\arcsec       & 50 sec & 120\arcsec$\times$150\arcsec   &  90 min.  &           & Intensity\\
%      & \cite*{Nov88}&           &        &            &  30 min.  & 4\arcsec $\times$ 4\arcsec    & Doppler  \\
%      &             &            &        &            & 190 min.  &           & velocity \\
%  \hline
 1989 &  Wang       &    1\arcsec       & 60 sec & 250\arcsec$\times$250\arcsec   &10 min. and&  3\arcsec $\times$ 3\arcsec   & Doppler  \\
      &\cite*{Wang89}&           &        &            & 60min.    &  9\arcsec $\times$ 9\arcsec   & velocity \\
  \hline
 1989 & November    &    1\arcsec       & 15 sec &  120\arcsec$\times$150\arcsec  &  90 min.  &  2\arcsec $\times$ 2\arcsec   & Intensity\\     
      &\cite*{Nov89a}&           &        &            &           &  4\arcsec $\times$ 4\arcsec   &          \\
  \hline
 1989 & Deubner     &     0.5\arcsec    &  6 sec &   224\arcsec      & duration= &           & Doppler  \\   
      &\cite*{Deubner89}&        &        &            &  32 min.  &           & velocity \\
  \hline
 1991 & Brandt et al.&   0.25\arcsec    & 12 sec &   14\arcsec$\times$12\arcsec   &  79 min.  & 0.4\arcsec$\times$ 0.4\arcsec & Intensity\\  
      & \cite*{BFSSTTT91}&       &        &            &           &           &          \\
  \hline
 1991 & Chou et al. &    2\arcsec       & 90 sec &            & k-$\omega$ &           & Doppler   \\  
      &\cite*{CLBD91}&           &        &            & analysis  &           & velocity \\
\hline
 1991 & Darvann     &   0.5\arcsec      & 15 sec &  150\arcsec$\times$120\arcsec  & 20 min.   &1.3\arcsec $\times$ 1.3\arcsec &Intensity \\
      &\cite*{Darv91}&           &        &            & 60 min.   & 5\arcsec  $\times$ 5\arcsec   &          \\
  \hline
 1992 &Straus et al.&   1\arcsec        & 94 sec &  240\arcsec$\times$120\arcsec  &   k-$\omega$     &           & Intensity\\
      &\cite*{SDF92} &           &        &            & analysis  &           & Doppler  \\
      &             &            &        &            &           &           & velocity \\
\hline
 1992 &Muller et al.&    0.25\arcsec    & 20 sec &   58\arcsec$\times$48\arcsec   &  17 min.  & 2.8\arcsec$\times$2.8\arcsec  & Intensity\\
      &        (1992)&           &        &            &           &           &          \\
\hline
 1992 & Chou et al. & 0.7\arcsec - 1.5\arcsec  & 60 sec &  136\arcsec$\times$100\arcsec  &   k-$\omega$     &           & Doppler  \\
      & \cite*{CCOW92}&          &        &  342\arcsec$\times$ 240\arcsec & analysis  &           &velocity  \\
\hline
 1993 &Abdussamatov&   0.6\arcsec      &   ?    &    138\arcsec     & spatial   &           &Intensity \\
      &\cite*{Abduss93}&         &        &    and     &Correlation&           &Doppler   \\
      &             &   0.25\arcsec     &1 Frame &  44\arcsec$\times$70\arcsec    & Int-Vit   &           &velocity  \\
\hline
 1995 & Wang et al. &    0.25\arcsec    & 20 sec &   31\arcsec$\times$31\arcsec   & 60 min.   &0.65\arcsec$\times$0.65\arcsec &Intensity \\
      & \cite*{WNTT95}&          &        &            &           &           &          \\
\hline 
 1997 & Straus and  &    0.5\arcsec     & 70 sec &   90\arcsec$\times$90\arcsec   &   k-$\omega$     &           &Intensity \\
      &Bonaccini \cite*{SB97}&   &        &            &analysis   &           &Doppler   \\
      &             &            &        &            &           &           &velocity  \\
\hline
 1997 &Bachmann     &    1.3\arcsec     & 60 sec &  200\arcsec$\times$200\arcsec  &  17 min.  &  (?)      &Doppler   \\
      &et al. \cite*{BKPH97}&    & (?)    &   (?)      &           &           &velocity  \\
\hline
 1998 &Roudier et al.& 0.25\arcsec -0.5\arcsec & 45 sec.&  60\arcsec$\times$65\arcsec    &  20 min.  & 3\arcsec$\times$3\arcsec      &Intensity \\  
      &\cite*{RMVP98}&           &        &            & 100 min.  & 5\arcsec$\times$5\arcsec      &          \\
      &             &            &        &            & 6h40min.  &           &          \\
\hline
 1998 & Ueno and    &  0.6-0.8\arcsec   & 15 sec.&  100\arcsec$\times$99\arcsec   &  20 min.  & 2.9\arcsec$\times$2.9\arcsec  &Intensity \\
      & Kitai \cite*{UK98}&      &        &            &  90 min.  &           &Doppler   \\
      &             &            &        &            &           &           &velocity  \\
\hline
\end{tabular}\end{flushleft}

\vspace*{5mm}
\caption[]{Mesogranulation measurements} \label{meso_table} 
\end{table*}

\clearpage} %end afterpage

Since its discovery by November et al. \cite*{NTGS81}, \MG\ has been
sought using
many different techniques (Doppler measurements, intensity variations,
horizontal velocity fields and their divergence) with the idea that one
could exhibit a quasi-steady cellular motion as clearly as granulation.
However, the reports of observations aimed at pointing out this new
feature of solar convection have never been clear-cut and always
difficult to compare to each other.

In Table~\ref{meso_table} we summarize all the previous studies on \MG. A
common result of these studies is that they always find some features in
the \MG\ range of scales, \ie within length scales between 4\arcsec\ and
12\arcsec\ (or 3~Mm and 10~Mm) and on time scales between 30~mn and 6~h.
The features are patterns either in intensity, 
or in radial velocity, or in horizontal divergence, for example.
The picture left by \MG\ observations is therefore fuzzy: neither
its characteristic size or time is well established and vary from
one author to another.

\begin{figure}[t]
\centerline{\psfig{figure=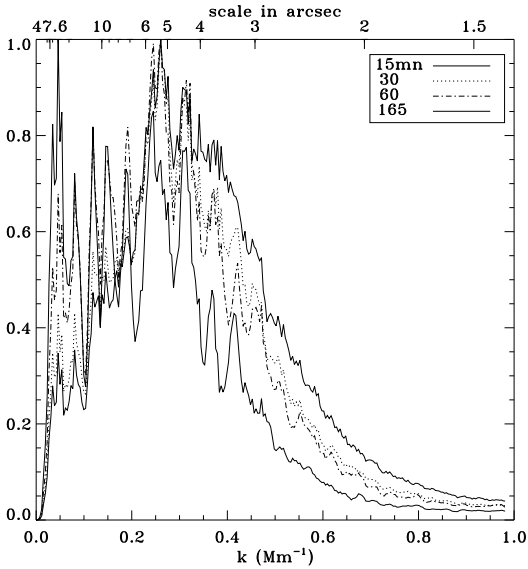,width=8cm}}
\caption[]{Power spectra $E_\tau(k)$ of the horizontal velocity fields
using the three hour data set of Pic du Midi. The grid size is
0\arcsec7 (1\arcsec = 728~km); note that the field of view is
58.2\arcsec$\times$47.6\arcsec.
All the spectra have been normalized by their maximum value.
The different line styles refer to the time-averaging window of size
$\tau$.}
\label{spect} \end{figure}

\begin{figure}[h]
\centerline{\psfig{figure=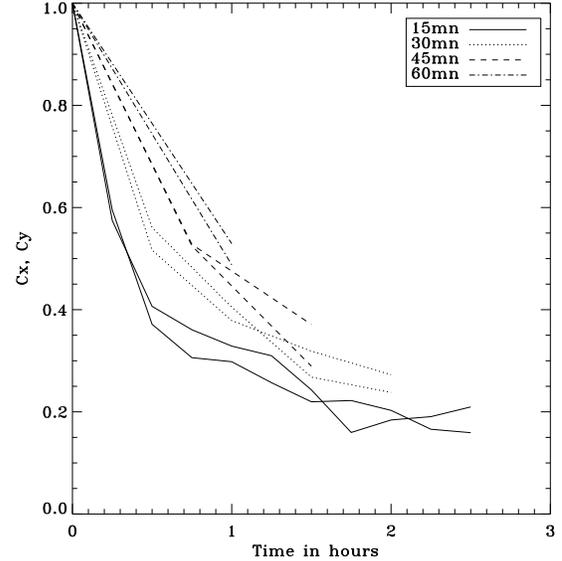,width=8cm}}
\caption[]{Linear correlation of successive time-averaged velocity fields. For
each averaging window we computed the correlation of both the $x$- and $y$-
components of the velocity field.}
\label{correl}
\end{figure}

\subsection{Spectra and correlations}

In order to better understand the situation, it is useful to consider the
physics which leads to the above mentioned observations. This is
obviously turbulent convection: motions seen at the sun's surface result
from the superposition of a large number of scales constituting the
turbulent spectrum. The relevant questions are therefore: why should a
scale like \MG\ single out among other scales? And if it does, how would
we recognize and characterize it?

Concerning granulation, the answers are known: it is the balance between
radiative diffusion and advection which determines the size of granules
and it emerges from other scales as the one with the highest contrast in
intensity and with the largest fluctuations of velocity. The spectrum 
of turbulent kinetic energy shows a maximum around this scale.

\begin{figure*}[ht]
\centerline{\psfig{figure=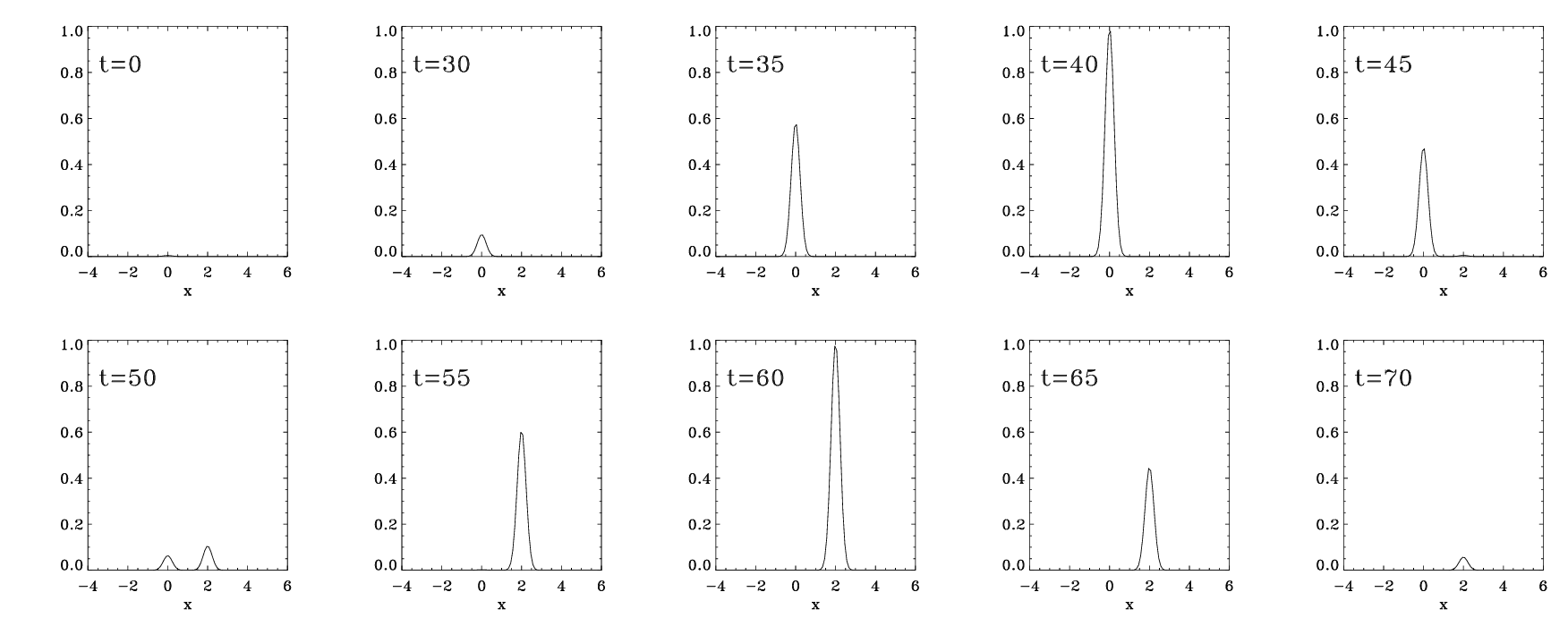,width=18cm,angle=90}}
\centerline{\psfig{figure=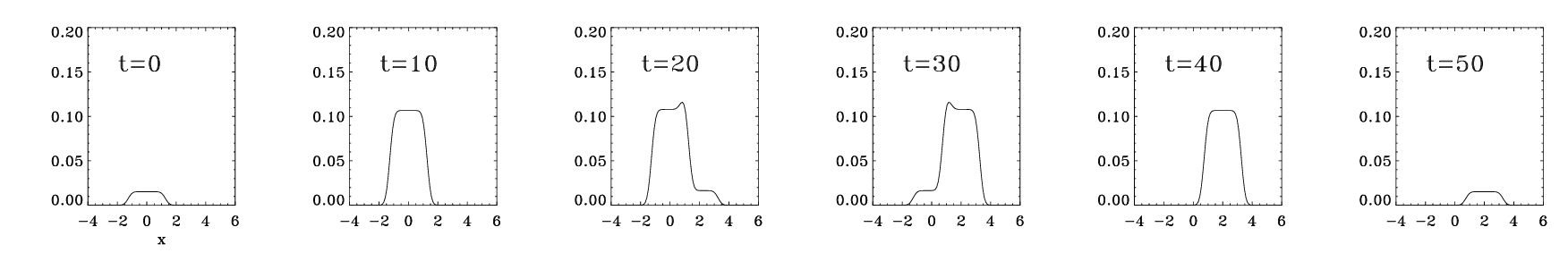,width=18cm,angle=90}}
\caption[]{First two rows: Time sequence of two structures (Gaussians) emerging
at two different places with a slight time-delay. Bottom row: The same
structures are now seen after convolution by a time window of width
$\Delta t=20$ and a spatial window of width $\Delta x=2.5$;
after a local growth, a single structure seems to move and then decays.}
\label{meso1}
\end{figure*}

At mesoscale, we do not know which mechanism could distinguish any particular
scale; we may, however, try to recognize a scale which stands out and
therefore search for a peak or a break of the slope in the kinetic energy
spectrum. Such a feature in the spectrum is indeed the signature that
some physical mechanism is injecting energy at a specific scale.

In previous work \cite{EMRN93}, spectra of kinetic energy have
not shown any
characteristic feature at scales larger than granulation. In fact,
large-scale features appear in turbulent flows with time averages.
Indeed, let us suppose
that the kinetic energy spectrum varies as $E(k)\propto k^{-\alpha}$ in
the mesoscale range; using dimensional arguments, it turns out that the
typical lifetime of structures of wavenumber $k$ is $\tau_k\propto
k^{(\alpha-3)/2}$ which means that the lifetime of turbulent structures
increases with their size since $\alpha <3$\footnote{The case
$\alpha=3$ corresponds to a two-dimensional turbulence; the turnover
time of eddies is then fixed by the background vorticity. In
three-dimensional turbulence, scales outside the dissipative range are
such that $\alpha\lesssim 5/3$.}. In other words, long
time-averages show large-scale features and the longer the average the
larger the scale. This point is clearly illustrated in Fig.~\ref{spect}
where we have computed kinetic energy spectra of the horizontal flow
derived from granule tracking in Pic du Midi data \cite{RRMV99}. These
spectra $E_\tau(k)$ are defined as follows: The mean kinetic energy (defined
by a time window of length $\tau$) at one point of the field reads:

\[ \demi\moy{v^2}_\tau = \int_0^\infty E_\tau(k)dk \]
where $E_\tau(k)$ is related to the Fourier transform of the velocity
components $\tilde{v}_x$, $\tilde{v}_y$ by

\[ E_\tau(k) = \int_0^{2\pi}\lp |\tilde{v}_x|^2+|\tilde{v}_y|^2\rp
kd\theta,\]
where $k_x=k\cth$ and $k_y=k\sth$ are the components of the wave vector.
In this
figure we clearly see the build-up of large-scales and the disappearance
of small scales when the time averaging window is made longer. The
average of the spectra of the velocity fields determined by LCT$_{\rm
bin}$ and a 15mn-window clearly shows a peak at a scale of
$\sim$5\arcsec\ (3500~km); this peak is the signature of
the most energetic horizontal flows.

\begin{figure}[ht]
\centerline{\psfig{figure=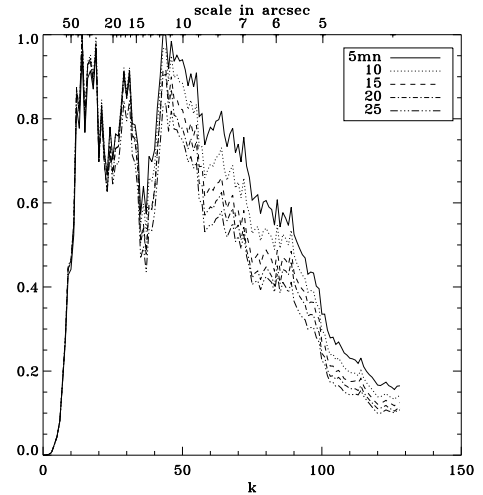,width=7cm}}
\caption[]{Same as Fig.~\ref{spect} but with a grid size of 1\arcsec96.}
\label{new_spec} \end{figure}

\begin{figure*}[t]
\centerline{\psfig{figure=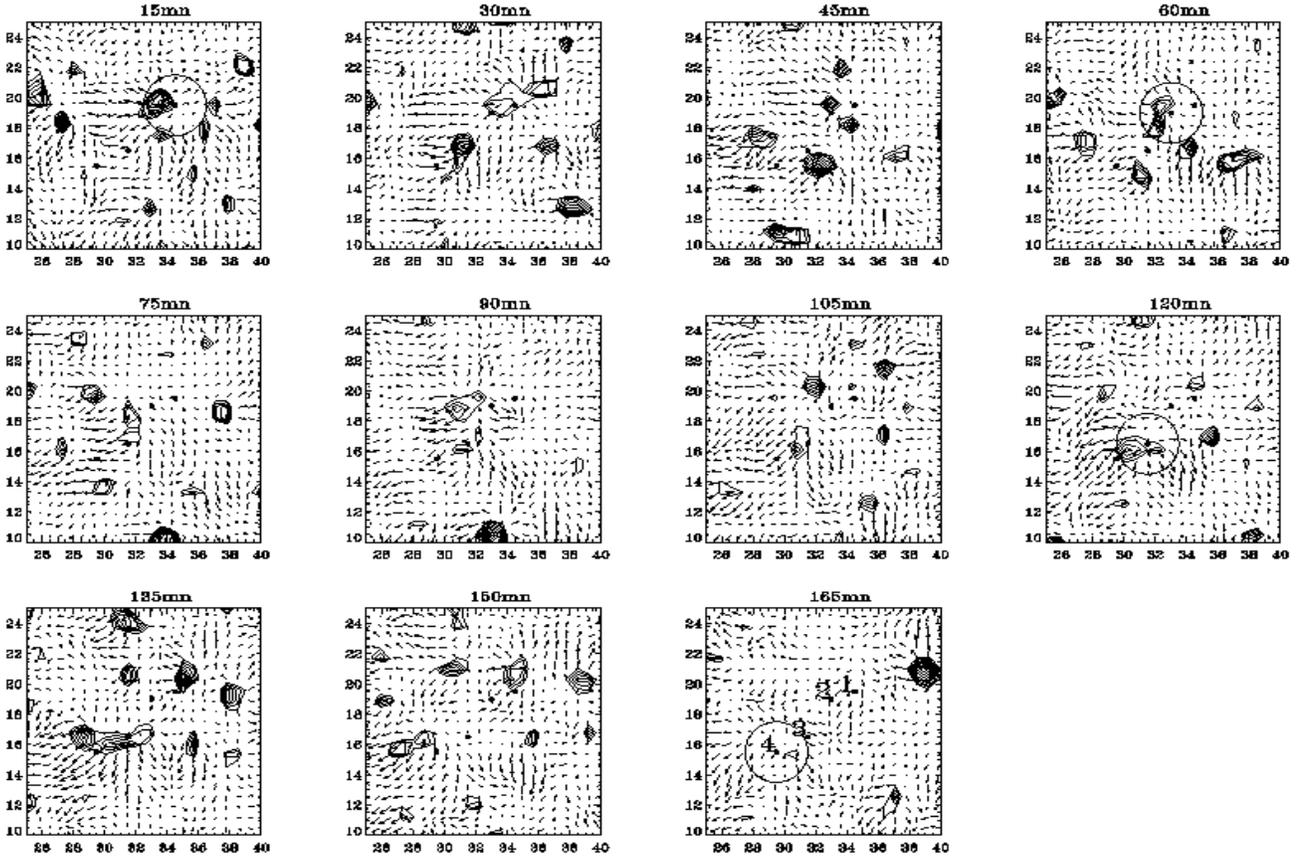,width=18cm,height=12cm,angle=90}}
\caption[]{Evolution of the flow field with the contours of high
positive divergence overlaid. The four black dots locate the position of
mesogranule 5 of Muller et al. \cite*{MARVSFST92} while the circle
emphasizes the position corresponding to the time of the snapshot. In
the last snapshot the numbers show the successive position of mesogranule
5. We clearly see from this sequence that no steady state structure of
the divergence field moves from position 1 to 4 during the sequence. Axis
units are in arcsec.}
\label{evol_meso} \end{figure*}

To further emphasize the turbulent nature of mesoscale features, we compute
the correlation between successive time-averaged velocity fields as a
function of time; we compute

\[ C_x = \frac{\overline{\lp\moy{v_x}_\tau-\overline{\moy{v_x}_\tau}\rp(t)
\lp\moy{v_x}_\tau-\overline{\moy{v_x}_\tau}\rp(t+n\tau)}}
{\sqrt{\overline{\lp\moy{v_x}_\tau-\overline{\moy{v_x}_\tau }\rp^2(t)}
\overline{\lp\moy{v_x}_\tau -\overline{\moy{v_x}_\tau }\rp^2(t+n\tau)}}} \]

\noi as a function of $n$ ($t$ is arbitrary); in this expression
overbars indicate spatial averages and $\moy{\cdot}_\tau$ stands for a
time average of length $\tau$. Results are plotted in Fig.~\ref{correl}.
They show that the autocorrelation of the mean velocity fields is halved
after one time step. This again emphasizes the role played by the time
window which selects a spatial structure whose lifetime is precisely of
the order of the time window, thus showing that no quasi-steady flow
exists on a time scale longer than 15~mn.

\subsection{Pitfalls of data processing}

Turbulence, however, is not the only difficulty of this problem; data
processing may also interfere and contribute to blur the results. 

As a first instance, let us compute the kinetic energy spectra of
horizontal flows, but using a larger spatial window (than in
Fig.~\ref{spect}) for the determination of the velocity field. The
result is shown in Fig.~\ref{new_spec} for a 2\arcsec-resolving window.
Very clearly the peak is now shifted to 12\arcsec\ (8.7~Mm). This
emphasizes the high sensitivity of the velocity field patterns (scales)
to the choice of the sampling window.

Finally, we would like to mention another effect of data processing
which can unduly extend the lifetime of mesoscale features. This is the
use of sliding time windows. Indeed, as illustrated in Fig.~\ref{meso1}
such windows can transform two independent time-evolving structures into
a single moving structure. This is what happened when Muller et al.
(1992) described a three-hour-living mesogranule which was used to show
the supergranular flow. In fact, as shown by Fig.~\ref{evol_meso}, an
independent time sampling shows that no coherent structure lasts such a long
time but new structures emerge after $\sim$30~mn.

We therefore interpret the results of previous work
on \MG\ as a consequence of uncontrolled averaging procedures.
This explains the large variability of results which have been published
on \MG: they are all dependent on the way authors have combined their
averages. 

\pagebreak

\begin{figure}[hbt]
\centerline{\includegraphics[width=6.5cm]{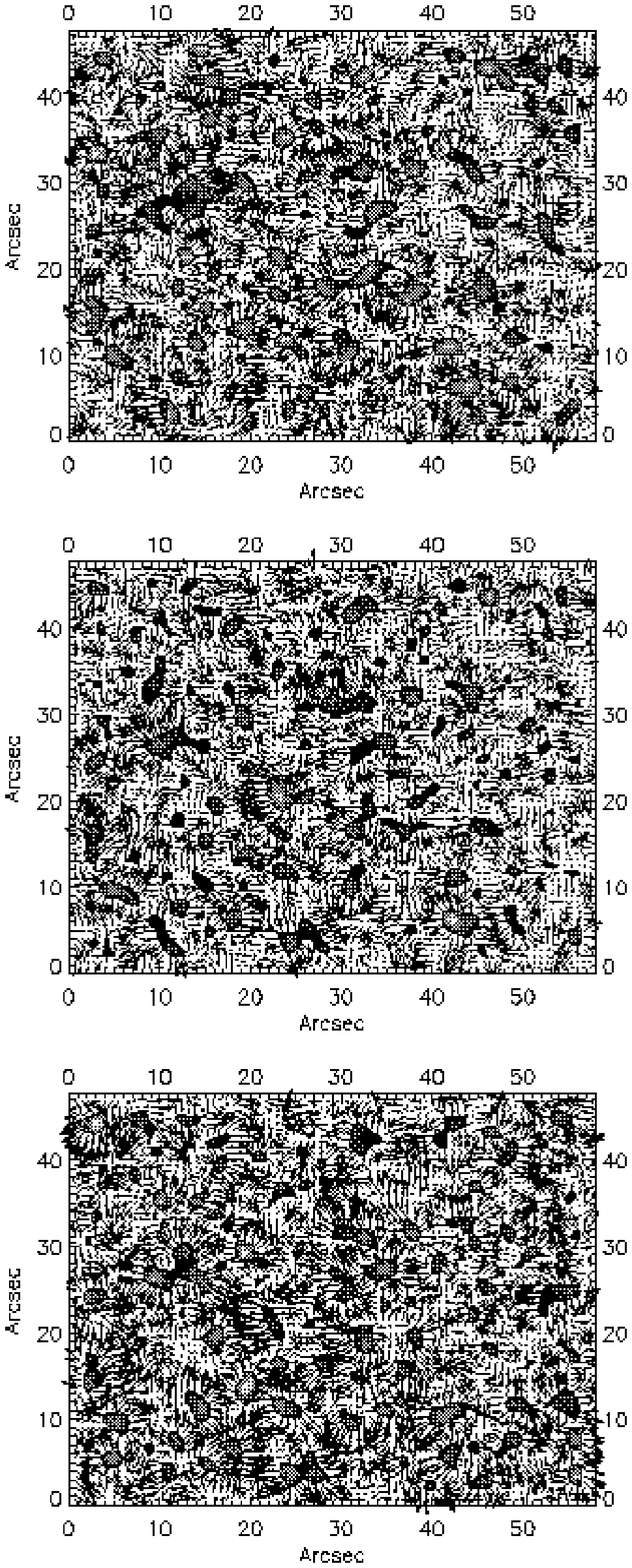}}
\caption[]{Evolution of the velocity field with the high positive values
(thresholded) of divergence overlaid. Each field was computed using a 5~mn
time-window with the LCT$_{bin}$ algorithm (cf Roudier et al. 1999).
The time step is 5~mn.} \label{evol_blob}
\centerline{\psfig{figure=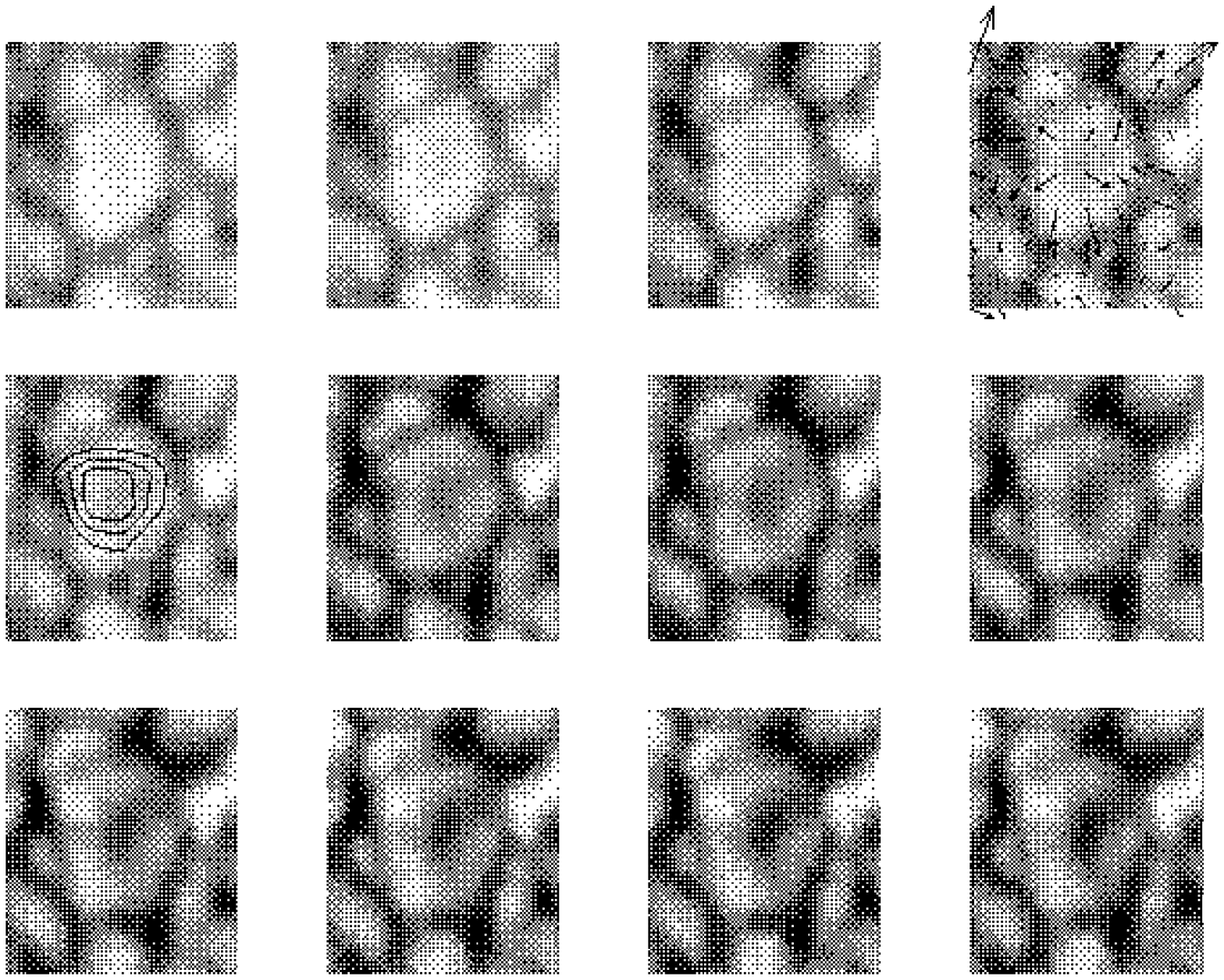,width=8cm}}
\caption[]{A time sequence of an exploding granule
with the corresponding velocity or divergence field overlaid. Each
snapshot is 5.6\arcsec wide and the time step is 20~s.}
\label{explos}
\vspace*{-2cm}
\end{figure}

\begin{figure}[t]
\centerline{\psfig{figure=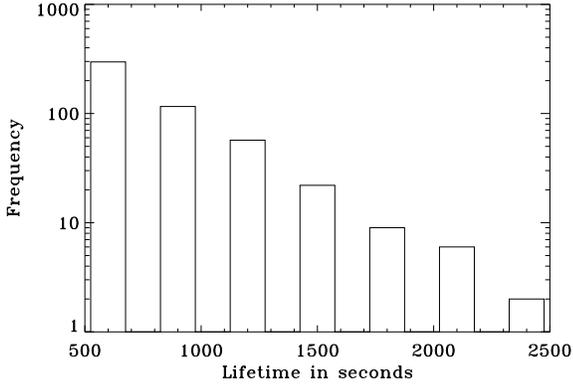,width=8cm}}
\caption[]{A histogram of the lifetime of patterns with positive
divergence above a given threshold. The lifetime is defined
as the time spent by a (simply connected) pattern over the threshold;
the mean lifetime is 15~mn.}
\label{histoblob}
\end{figure}

\subsection{Strong positive divergences (SPDs)}

The foregoing results, however, leave unanswered the question of the
origin of kinetic energy in the mesoscale range; we may indeed wonder
which flow patterns are contributing to the spectral peak at
$\sim5$\arcsec\ in Fig.~\ref{spect}. A plot of the velocity field along
with its divergence (Fig.~\ref{evol_blob}) shows that the strong
horizontal flows are generally associated with strong positive
divergences (SPDs) among which the exploding granules are the most
energetic. This is well illustrated by the two time sequences in
Fig.~\ref{evol_blob} and Fig.~\ref{explos}. From
Fig.~\ref{evol_blob}, it is quite clear that the divergence field is
highly variable, showing patterns which may extend up to
10\arcsec\ (7.3~Mm). A histogram of the lifetime of these patterns
(Fig.~\ref{histoblob}) gives a mean lifetime of 15~mn which is short.

Ending this section, we are therefore lead to the conclusion
that no specific scale exists in the \MG\ range
except the scale of horizontal flows featured by SPDs.  Hence we
confirm the conclusion of Straus and Bonaccini \cite*{SB97} that no gap
in the kinematic energy spectrum separates granulation from \MG\ which
therefore must be considered just as the large-scale extension of
granulation.

\section{Supergranulation}

\subsection{Mean flows and SPDs}

\begin{figure}[t]
\centerline{\psfig{figure=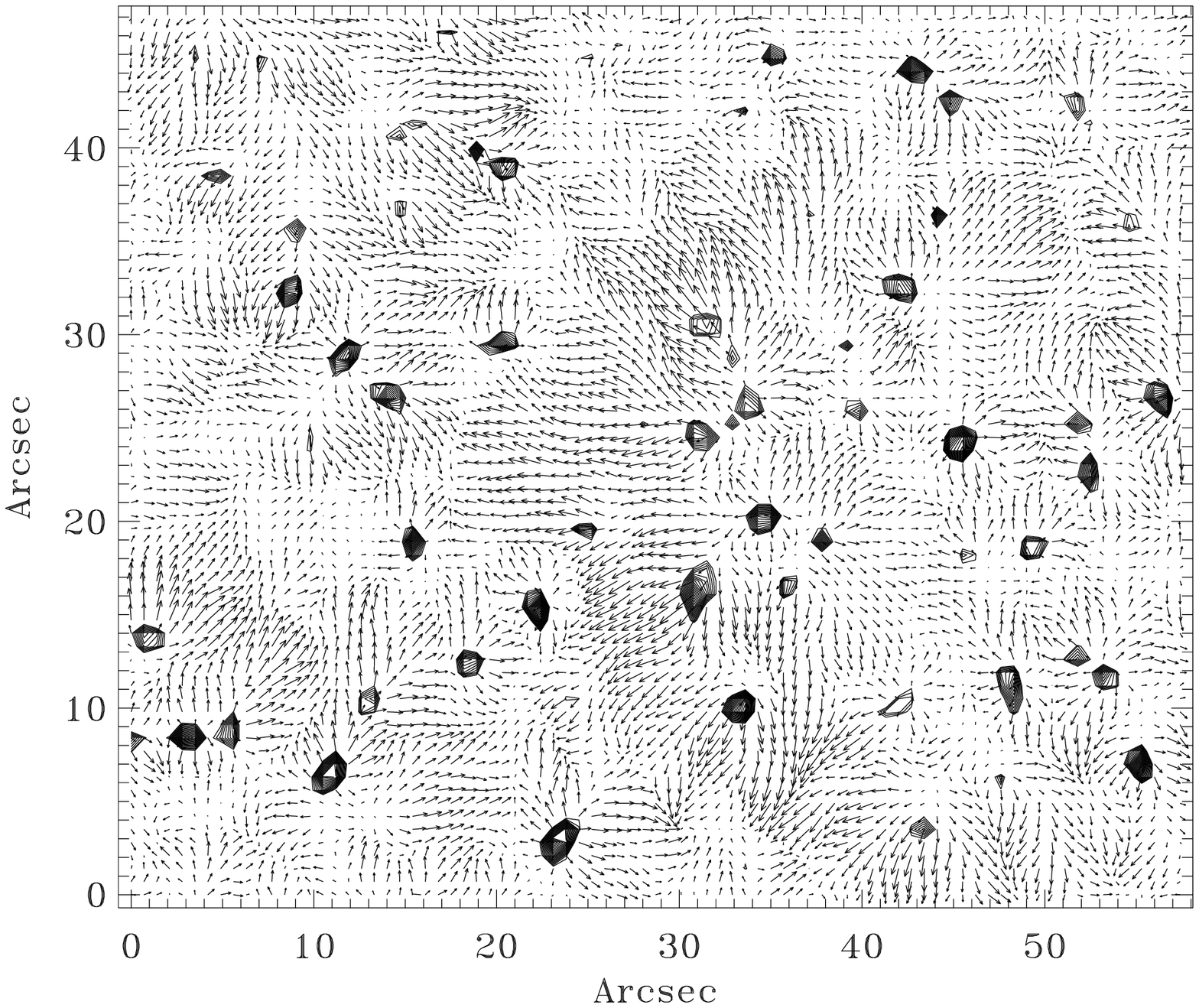,width=9.4cm,height=6.5cm}}
\centerline{\psfig{figure=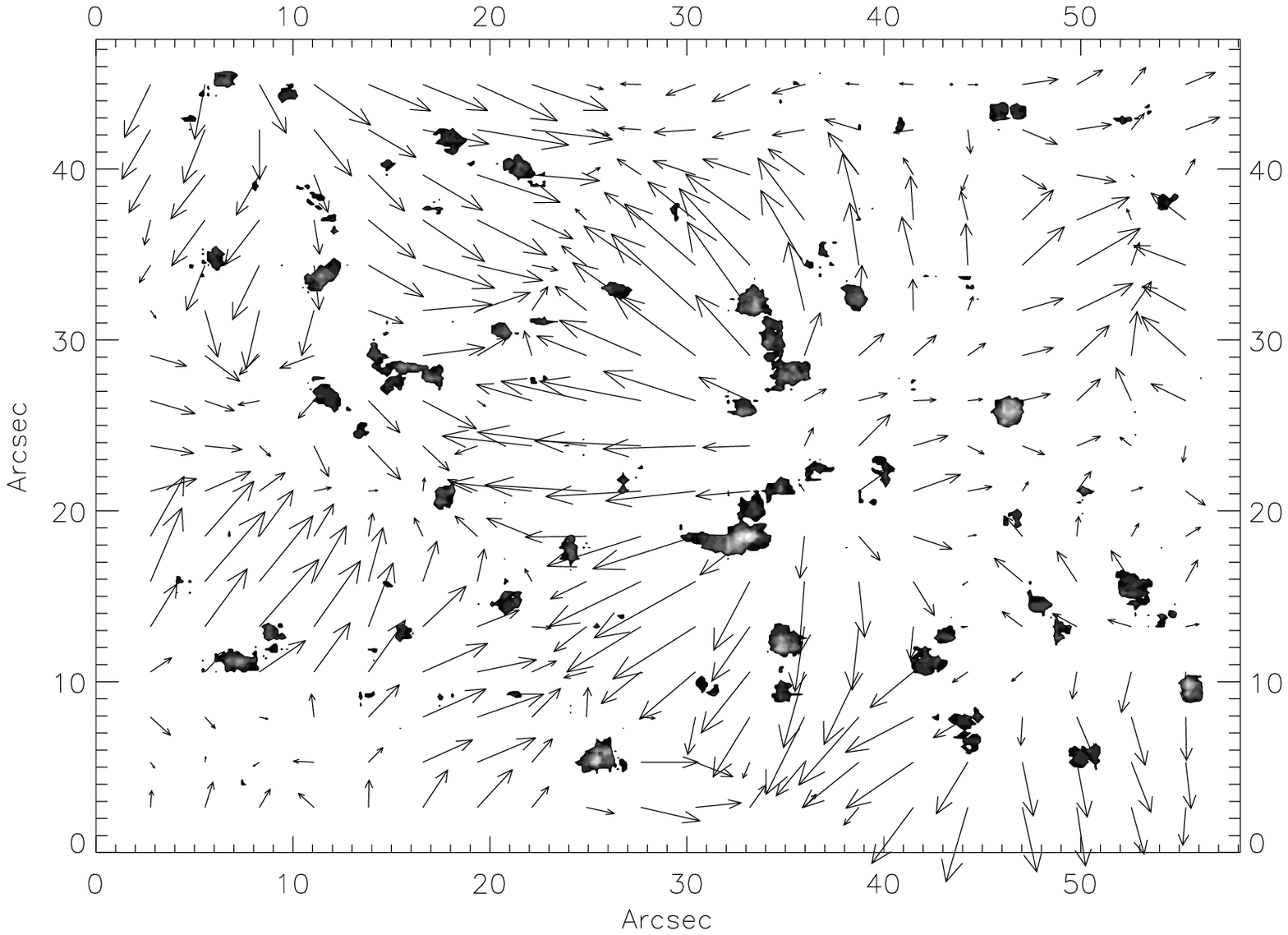,width=9cm}}
\caption[]{The three hour time average of the velocity field
using a 0.7\arcsec window (above) and a 2.8\arcsec window (below).
Contours show the thresholded positive divergence.}
\label{mean_blob}
\end{figure}

We may now wish to know whether \SG\ plays some part in the motion
of granules. A first look at the spectrum of the 3-hour-average
velocity field (Fig.~\ref{spect}) shows some energy at $\sim30$\arcsec\
(22~Mm). However, this spectral peak is only suggestive since 
the field of view  is just 47.6\arcsec. Turning to real space (as
opposed to spectral space), a plot of the velocity field 
(Fig.~\ref{mean_blob}) shows
that indeed a large-scale velocity field may exist. The rms velocity of
the 3-hour-average field is $\sim230$~m/s. On the unfiltered field we
clearly see the imprint of SPDs with their typical size of 5\arcsec; this
indicates that the origin of the mean flow may be found in the
cumulative effects of SPDs. Furthermore, if this mean field is
identified with \SG, we have at hand its origin: correlated SPDs.

\subsection{Network formation from measured velocity fields}

\begin{figure}[t]
\centerline{\psfig{figure=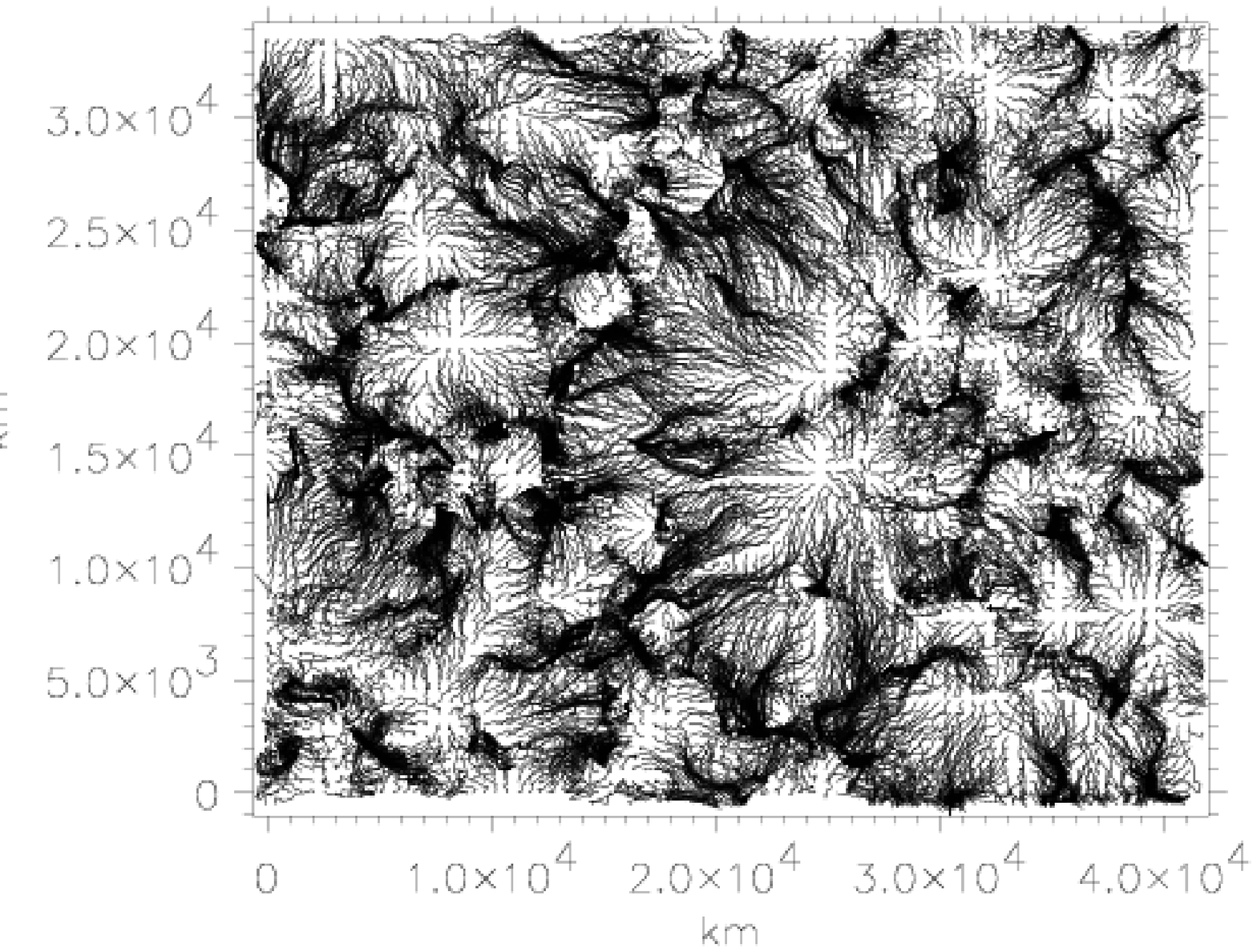,width=8.2cm}}
\vspace*{-2.5mm}
\centerline{\psfig{figure=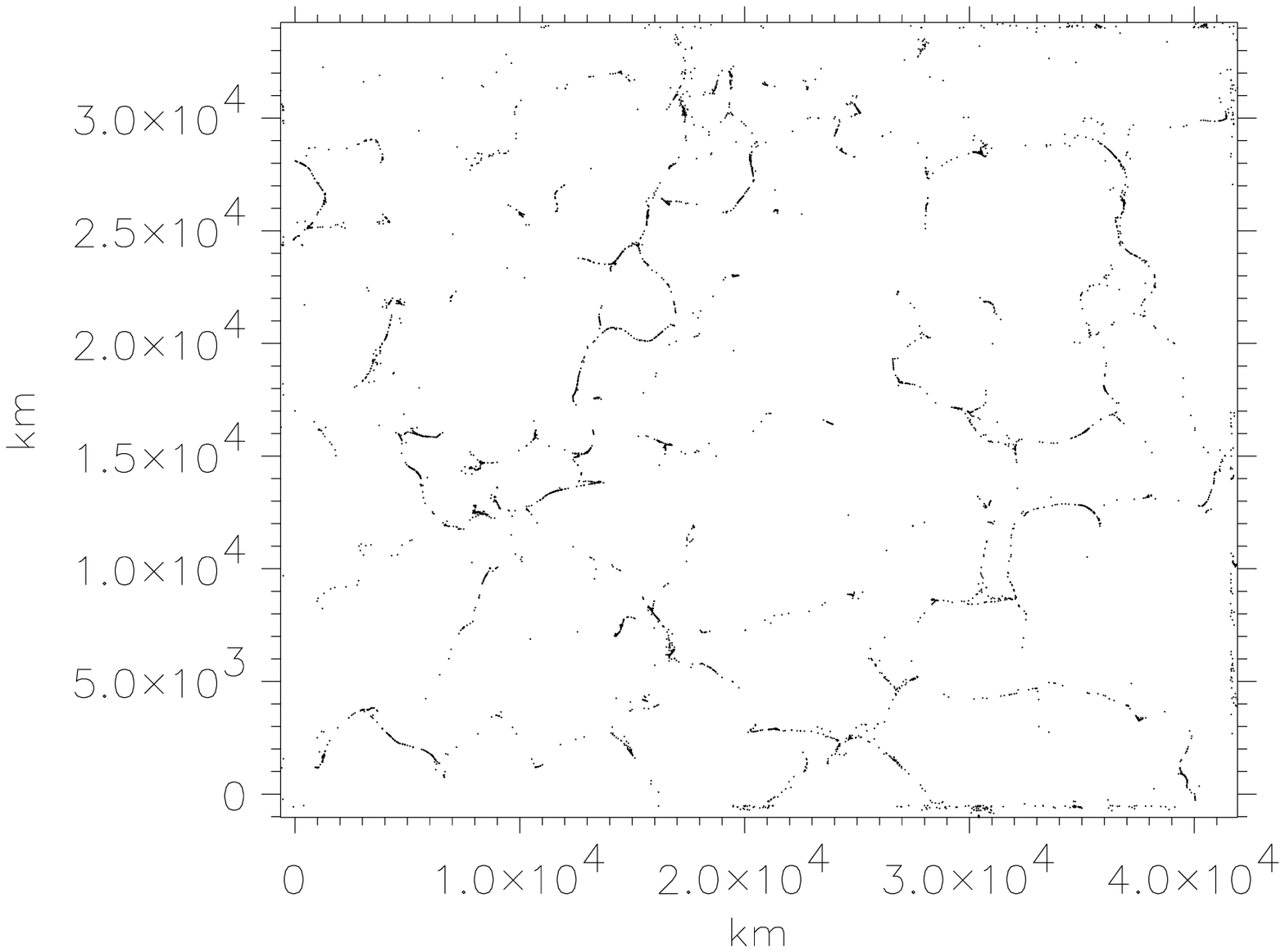,width=8.2cm}}
\vspace*{-2.5mm}
\centerline{\psfig{figure=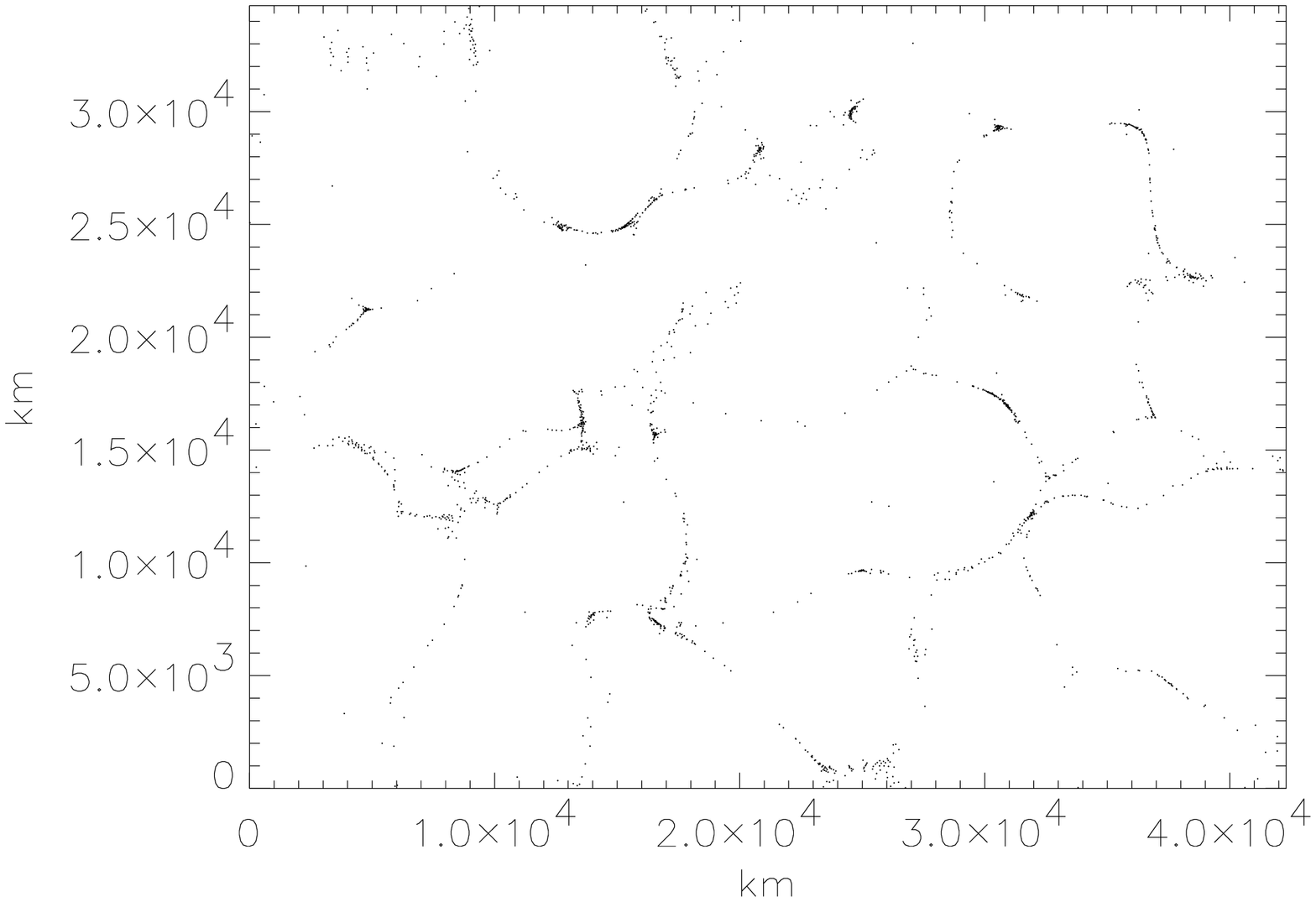,height=6.5cm,width=8.2cm,angle=90}}
\caption[]{Top: Corks' trajectories for the velocity field derived
from granule motions in the Pic du Midi data set. Middle: Corks'
distribution after three hours of advection. Bottom: Same as (middle)
but with simulated velocity fields (see text).}
\label{cork_traj}
\end{figure}

One way to confirm the view that a three-hour time average of
horizontal flows shows the supergranulation scale, is to focus on
its transport properties. This technique was
already used in the past by Simon and Weiss \cite*{SW89}. It consists of
integrating the trajectories of floating corks, initially uniformly
distributed, and characterizing their spatial distribution after some
time. Simon and Weiss used the SOUP data sequence, which lasted 28~mn,
to determine a kinematic model of the horizontal velocity field. Then,
they integrated the cork trajectories during a time very much longer (up to
16~h) than the data sequence. Our method is much closer to the
data: with the techniques described in Roudier et al. (1999), we
determine the velocity field evolution of the Pic du Midi data set with
a high spatial and temporal resolution (0\arcsec7 and 5~mn). We then use
this field to integrate the cork trajectories during the three hours of
recorded data.

\begin{figure}[t]
\centerline{\psfig{figure=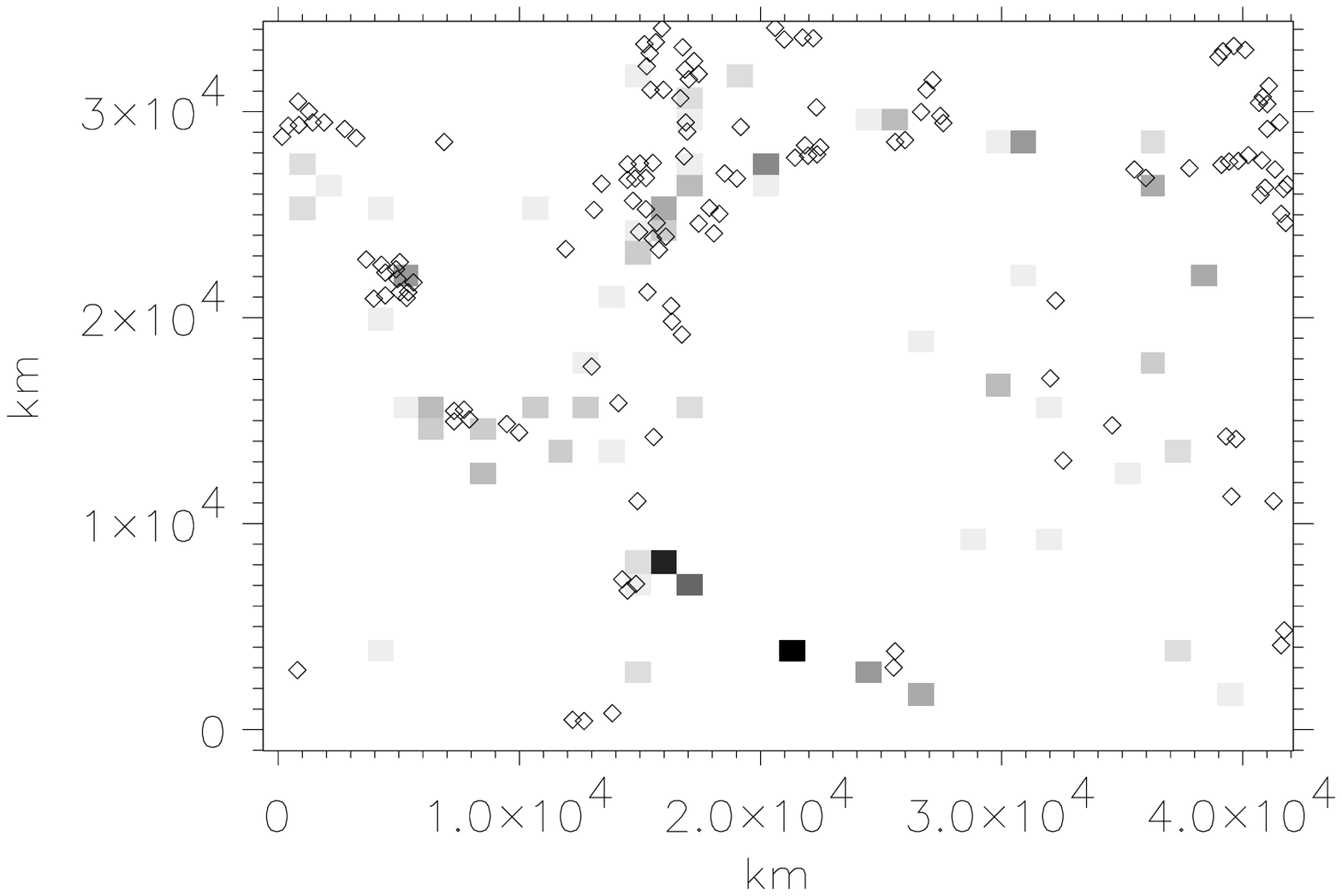,width=9cm}}
\caption[]{Corks' density (squares in grey scale) and network bright points
in the same field of view; their position are derived from Muller et al.
1992. The distribution shown by the squares represents 70\% of the
corks, the rest being distributed randomly in the field.}
\label{NBP}
\end{figure}

As shown in Fig.~\ref{cork_traj}, corks are expelled from certain 
regions, whose
size varies between 5~Mm and 15~Mm and again SPDs seem to play a
major role. We cannot say that the cells thus formed are
``supergranules" since they would continue to evolve if our data set
were longer. If the concentration of corks is plotted, however, the
supergranulation scale already appears. The clearest evidence of
this is given by the remarkable coincidence between the regions with a high
density of corks and the positions of network bright points
(Fig.~\ref{NBP}). This result should be compared with the one of Brandt
et al. \cite*{BRST94}
who found a similarity between cork distribution and the
Ca-network.

Referring back to the mean divergence field (Fig.~\ref{mean_blob}),
we also see that the regions with
negative divergence are the ones where corks and network bright
points tend to concentrate.

Finally, we also used corks to estimate a horizontal diffusivity
$D=<r^2\!>/4t$ following Berger et al. \cite*{BLST98}. We found values
between 50~km$^2$/s and 100~km$^2$/s but, as this diffusivity did not
reach an asymptotic value at the end of our time-sequence, we think
that the aforementioned values only show the amplitude of the transport
but nothing about its physical origin (advection, diffusion, abnormal or
turbulent diffusion).

\subsection{Network formation from simulated velocity fields}

In order to show the leading role of SPDs, we extracted from the data
their positions in space and time as well as their
mean radii. We used them to determine the amplitudes ($V_n$) of the
model flow used by Simon and Weiss (1989) and Simon et al.
\cite*{STW91}, namely

\[ \vv = \sum_n V_n {|\vr-\vr_n|\over R}\; e^{-|\vr-\vr_n|^2/R^2}\en \]
where $\vr_n$ designates the position of SPDs, $R$ their range
of influence, which lies between 0.5\arcsec\ and 3\arcsec\ with a mean
value of 2.3\arcsec, and $\en$ the direction of their flow.

Using such a model flow, we have repeated the calculation for the
advection of corks and found a distribution very similar to the one
obtained when using the full data (see Fig.~\ref{cork_traj}).

It therefore turns out that SPDs give the main contribution to the
horizontal transport of passive scalars.

\section{Discussion}

\subsection{Mesoscale flows}

The analysis of flow motions at mesoscale that we presented in Sect.~3
has shown that the scale of flow patterns is controlled by the size of
the averaging time window. We interpret this result as evidence of
the turbulent nature of the kinetic energy spectrum in this range of
scales. We therefore refute the idea that some quasi-steady flow exists
at mesoscale and show that previous identifications of mesogranules
advected by a supergranular flow, as shown by Muller et al. (1992) for
instance, are an artefact of the averaging procedure. We identify the
only mechanism able to structure the horizontal flow field as being
SPDs among which we find exploding granules. 
However, the scale controlled by SPDs is
rather centered around 5\arcsec\ (3.5~Mm) which is small compared to the
scale generally attributed to mesogranulation (5-10~Mm); we conjecture
that these latter scales are in fact built up by the nonlinear
interactions of the former ones and form the (turbulent) continuation of
the granular motion spectrum. Using two-dimensional simulations Ploner
\cite*{Ploner98} also concludes that mesoscale flows can result from
nonlinear interactions of granular flows.

We have shown subsequently that the use of Pic du Midi velocity fields in
computing trajectories of floating corks, shows that at the end of the
time integration, which we take equal to the length of the data set,
corks concentrate in regions which coincide with those occupied by
network bright points.
It thus turns out that in a rather short time the
supergranulation scale emerges from the flow delineated by granule
motions.

These results show that the traditional view which assumes
supergranulation as a quasi-steady horizontal flow driven by the
ionization of helium is certainly too simple. Obviously the build-up of
the supergranulation scale as far as magnetic flux tubes are concerned,
results from a process of turbulent transport in which SPDs play an
important part.

\subsection{A model for supergranulation}

In order to explain the above mentioned observations and some others, we
now present a model which seems to square with all the known constraints
of \SG.

Let us consider the surface convection of the sun as a set of
granules. Assume that each granule interacts nonlinearly, mainly with its
nearest neighbours. The set of granules behave like a set of
nonlinearly coupled oscillators. A general behaviour of such a system is
that its energy can be focused on one or very few oscillators whose
motion is then very much enhanced \cite{DP93}. Such oscillators
collecting the
kinetic energy of the others would appear, in the solar context, as
SPDs or even exploders. This may also be seen as a manifestation of
intermittency of the turbulent solar flow. Now it has been noticed that
such exploding granules appear generally in neighbouring places
(see the mean divergences) which means that some temporal and spatial
correlations exist between them, namely that a large-scale flow organizes
their appearance (or vice-versa). This large-scale flow is obviously
\SG; its origin should be found in a large-scale instability of the
flow expressed by the set of granules.

Recent theoretical work \cite{GVF94,SSSF89} on turbulent flows has
indeed shown that a large-scale perturbation of some given steady
spatially periodic flow can be unstable in some bandwidth of wavenumbers.
Typically two kinds of situations may occur: the original small-scale
flow is invariant with respect to parity or not. If it is not,
large-scale instabilities occur through an AKA\footnote{Anisotropic
Kinematic Alpha effect: it is the Reynolds stress dependence with
respect to the mean velocity field; it occurs when the turbulence has
helicity, \ie lacks parity invariance.} effect which is the
equivalent of the $\alpha$-effect of turbulent MHD flows; if it is parity
invariant, which is the case of solar granulation (flows are hardly
helical), then instability appears through a negative eddy viscosity; a
range of large-scale modes is then destabilized and some large-scale
coherent flow starts. However, this new flow is rapidly hindered by the
motions it induces: as the Reynolds stress distribution is modified,
usually the turbulent viscosity comes back to positive values; in
our case we are considering turbulent convection and any flow increasing the
convective heat flux will be slowed by the overcooled material.
A similar scenario was also envisaged by Krishan \cite*{Krish91} using
arguments based on inverse cascade of two-dimensional turbulence;
indeed, large-scale instabilities are a way of realizing an inverse
cascade.
However, the mechanism invoked by Krishan is rather close to the
AKA-effect which requires helical motions at supergranular scales; such
motions do not seem to be observed. On the other hand, higher in the
atmosphere, the stratification stabilizes the fluid and may produce
two-dimensional motions which can ease energy transfer to large scales.

We therefore understand why large-scale (with respect to granulation
scale) flows like \SG\ will appear. We also understand why \SG\ does
not realize a
perfect pavement of the whole sun as does granulation. The development
of the large-scale instability is a stochastic process which results
from a given arrangement of the granular flow. It has not the constraint, as
any convective flow, to transport heat.

We also understand why the main downflows of a supergranule are not at
its boundary as revealed by the MDI instrument on board of SOHO (see
http://sohowww.nascom.nasa.gov/gallery/MDI/ mdi009.html,mdi009.ps or Zahn
1999\nocite{Zahn99}). They are in fact associated with the most
energetic granules \ie the exploders, and therefore occur mainly in the
bulk of the supergranule. The downflows on the boundaries should not be
very intense since the supergranular flow {\em must remain weak}
compared to the granular flows.

Hence, this model explains an interesting number of observations.
Essentially, it states that supergranular flows are surface flows which
do not result from a large-scale thermal instability.

\section{Conclusion}

To conclude this paper we would like to emphasize the point that the
picture of surface solar convection which has been popular in the
literature and where supergranulation advects \MG\ which in turn advects
granulation is too simple and misleading. It is obviously too simple
because it tentatively describes turbulent convection with three
scales instead of a continuous spectrum of scales. It is misleading as
it reduces the nonlinear interaction between scales to a simple
advection as for instance the kinematic models of Simon et al. 1991.

We have shown here that no quasi-steady flow could be identified at the \MG\
scale and that after a three hour averaging the mean flow shows a
component at the \SG\ scale while it keeps a small-scale (5\arcsec)
component. This latter component seems to be the source of the former
large-scale flow.

Our results therefore suggest a scenario where the large-scale
supergranular flow is generated directly by the granular flow through a
large-scale instability which fixes the scale, in space and time, of
\SG. We thus conjecture that nonlinear interaction between flows at the
granulation scale, in other words Reynolds stresses, are sufficient to
drive flows at the supergranulation scale and that the energy released by
the recombination of ionized helium plays no part. This scenario needs
now to be tested for its various implications, theoretical as well as
observational.

\begin{acknowledgements}
We would like to thank very much Peter Brandt for very fruitful and
helpful comments on the manuscript. We also appreciated the 
discussions and comments of our colleagues Keith Aldridge, G\'erard
Coupinot, Richard Muller and Sylvie Roques.  We are very grateful to
Richard Muller for providing us with the three-hour sequence of Pic du
Midi.  Special thanks are due to the Pic du Midi Observatory staff for
their technical assistance.

\end{acknowledgements}

\bibliography{}

\end{document}